\documentclass{iopart}
\usepackage{verbatim,hyperref}
\usepackage[english]{babel}

\newcommand{\llsim}{\stackrel{<}{\sim}}
\begin{document}

\title[Exact solutions of relativistic
magnetohydrodynamics equations]{Exact Solutions of the
Self-consistent System of Relativistic Magnetohydrodynamics
Equations for an Anisotropic Plasma on the Background of
Bondi-Pirani-Robinson's Metric}
\author{Yurii Ignatyev, Alexander Agathonov}

\address{Department of Mathematics, Kazan (Volga Region) Federal University, Mezhlauk 1
str., Kazan 420021, Russia}

\ead{ignatev\_yu@rambler.ru, a.a.agathonov@gmail.com}

\begin{abstract}
Exact solutions of the self-consistent relativistic
magnetohydrodynamics equations for an anisotropic magnetized plasma
on the background of Bondi-Pirani-Robinson's vacuum plane
gravitational wave (PGW) metric with an arbitrary polarization are
obtained, which generalize the results obtained earlier by one of
the authors for the transverse polarization of a gravitational wave.
Based on the reformulated energobalance equation it is shown that in
the linear approximation by gravitational wave amplitude only the
transverse polarization of PGW interacts with magnetized plasma.
\end{abstract}


\maketitle

\section{Introduction}
The equations of the relativistic magnetohydrodynamics (RMHD) of a
magnetoactive plasma in a gravitational field were formulated in
\cite{Ign95}\footnote{Before 2000 Yu.G. Ignatyev wrote his name as
Yu.G. Ignat'ev.} using the equality requirements for the dynamic
velocities of the plasma and the electromagnetic field\footnote{This
requirement is completely equivalent to the condition of plasma
infinite conductivity, see Ref.\cite{Ign95}.}. These equations were
obtained on the basis of the Einstein and Maxwell equations. A
remarkable class of {\it exact solutions} of these RMHD equations
was also found. It explains the motion of a magnetoactive locally
isotropic plasma in the field of a plane gravitational wave (PGW).
This class was called {\it gravi\-magnetic shock waves} (GMSW). It
describes {\it essentially nonlinear processes} which do not exist
in the linear approximation of magnetohydrodynamics and essentially
relativistic processes in terms of predominance of the massless
electromagnetic component in the magnetoactive plasma.
It was shown in \cite{Ign96} that the GMSW in pulsar magnetospheres
may be the highly effective detectors of gravitational waves from
neutron stars. Particularly, giant pulses which sporadically appear
in the radiation of some pulsars may be observational results of
energy transfer from a gravitational wave to GMSW. Estimations made
in \cite{Ign96}-\cite{IgnPhys97} make it possible to connect giant
pulses in radiation of the pulsar B0531+21 with gravitational
radiation in the basic mode of oscillations from this pulsar. In
fact, at present it is rather difficult to speak of identification
of giant pulses as an electromagnetic display of gravi\-magnetic
shock wave evolution in the pulsar magnetosphere and to
unambiguously connect these pulses with the pulsar's gravitational
radiation. Nevertheless, the idea of analyzing the effect of
gravitational waves from a compact astrophysical object on its own
electromagnetic radiation is highly productive for solving the
problems of gravitational waves detection.
In fact, the main difficulties of gravitational waves detection in
the Earth environment are:
\begin{enumerate}
\item An extremely small amplitude of gravitational waves on Earth
($h \llsim 10^{-19}$) due to significant distances from relativistic
astrophysical objects.

\item A sporadic nature of events leading to radiation of
gravitational waves inside relativistic astrophysical objects with
sufficient power. This does not allow one to unambiguously connect a
received signal with a fact of gravitation radiation detection.

\item Impossibility to construct relativistic detectors with
anomalous, highly effective parameters for gravitational wave
detection in conditions terrestrial laboratory (superstrong magnetic
fields, a highly anisotropic working body of the detector, low level
of background noise etc.).
\end{enumerate}
It is possible to avoid these problems if one could transfer a
detector directly to a close neighborhood of a relativistic
astrophysical object. In this case, one always has a ready
electromagnetic signal and there is no need to convert it to other
forms, which allows for conducting correlation analysis. If the
working body of the detector is the magnetosphere of a relativistic
astrophysical object, the optimal parameters for gravitational wave
detection are achieved automatically: super-strong magnetic fields,
an ultrarelativistic equation of state, highly anisotropy etc.
The fundamental importance of the GMSW for gravitational theory, as
a direct conversion effect of gravitational wave energy into
electromagnetic energy, leads to a necessity of a more detailed and
comprehensive researches. In \cite{IgnKin97}, a strict foundation of
the GMSW hydrodynamic theory was formulated on the basis of the
relativistic kinetic theory. As has been shown in
\cite{Ign95}-\cite{IgnPhys97}, a GMSW is realized in an essentially
collisionless nonequilibrium plasma in anomalously strong magnetic
fields. The isotropy of a local plasma electron distribution
essentially is violated under such conditions due to strong
bremsstrahlung. Therefore, the anisotropy factor of a magnetoactive
plasma is highly essential for the efficiency of the GMSW formation
mechanism. A hydrodynamic model of GMSW in an anisotropic plasma was
constructed in \cite{IgnGor97} for a specified relation between
parallel and perpendicular components of the plasma pressure. The
analysis in \cite{IgnGor97} was based on the general RMHD equations.
Particularly, an elementary linear relation was considered in
\cite{IgnGor97}. This study has revealed a strong dependence of the
GMSW effect on the plasma degree of anisotropy. That fact has led to
a necessity of constructing a dynamic model for the motion of an
anisotropic magnetoactive plasma in the field of gravitational
radiation.

Further, in \cite{IgnAga10} a detailed numerical model of GMSW has
been carried out at various parameters of the anisotropic
magnetoactive plasma in the computer algebra system Mathematica.
This study was based on the numerical solution of the nonlinear
energybalance equation by means of special methods of numerical
integration. The results received in \cite{IgnAga10} have confirmed
earlier made analytical estimations of the magnetoactive plasmas
behavior in a field of strong gravitational wave and have defined
more exactly some characteristics of GMSW.

However, at all variety of models of plasma only the case of
monopolarized gravitational wave with polarization $\mathbf{e}_+$
investigated in all quoted papers. It has been thus shown that the
case of monopolarized gravitational wave with polarization
$\mathbf{e}_\times$ is reduced to the case of $\mathbf{e}_+$
polarization state at the coordinate and physical quantities
transformations. However, the case of simultaneously existence of
two polarizations of a gravitational wave was not investigated. In
this paper we consider such a case. Thus it was possible to reduce
some additional conditions, which in \cite{Ign95} have defined the
structure of potential electromagnetic field in a magnetized plasma,
and thereby give a more general meaning to results obtained in
preceding papers. In this paper adopted a system of units where $(c
= G = \hbar = 1)$.

\section{Self-consistent RMHD equations in a gravitational
field}
\subsection{Frozen-in condition of magnetic field in plasma}
In \cite{Ign95} under the assumption of equality of dynamic timelike
velocity $v^i$ of a plasma and electromagnetic field\footnote{The
index ``$p$'' refers to the plasma, the index ``$f$'' to the field,
a comma denote covariant derivatives.}:
\begin{equation}\label{eq_vel}
\stackrel{p}{T}_{ij}v^j=\varepsilon_p v_i;\quad \stackrel{f}{T}
_{ij}v^j=\varepsilon_f v_i;\quad (v,v)=1
\end{equation}
on the basis of conservation of the total energy-momentum tensor of
a plasma and electromagnetic field,
\begin{equation}\label{TEM}
T^{ij}=\stackrel{p}{T}\ \!\!\!^{ij} + \stackrel{f}{T}\ \!\!\!^{ij},
\end{equation}
\begin{equation}\label{conserv_TEM}
T^{ij}_{~,j}=0
\end{equation}
full self-consistent system of relativistic magnetohydrodynamic
equations for magnetized plasma in arbitrary gravitational field has
been obtained. It describes the motion of a relativistic plasma and
an electromagnetic field in the given gravitational field.

In particular, it was shown that at positivity of the first
invariant of an electromagnetic field:
\begin{equation}\label{I_inv}
{\rm Inv}_1=F_{ij}F^{ij}=2H^2>0
\end{equation}
and equality to zero of the second invariant:
\begin{equation}\label{II_inv}
{\rm Inv}_2=\stackrel{*}{F}_{ij}F^{ij}=0
\end{equation}
necessary and sufficient condition for the solvability of equations
(\ref{eq_vel}) is {\it the frozen-in condition of magnetic field in
plasma}, i.e., equality to zero of the accompanying intensity of the
electric field $E_i$:
\begin{equation}\label{embed}
E_i= F_{ji}v^j=0.
\end{equation}
In formulas (\ref{I_inv})-(\ref{embed}) and further: $F_{ij}$ -
antisymmetric Maxwell tensor, $\stackrel{*}{F}_{ij}$ - dual Maxwell
tensor.
\begin{equation}\label{F_dual}
\stackrel{*}{F}_{ij}=\frac{1}{2}\eta_{ijkl}F^{kl},
\end{equation}
where $\eta_{kilm}$ - covariantly constant discriminant tensor
(Levi-Civita's tensor) \cite{Sing}.

At the frozen-in condition (\ref{embed}) fulfilment, the
condition of dynamic velocities equality (\ref{eq_vel}) is always
satisfied, regardless of the conditions
(\ref{I_inv})-(\ref{II_inv}). On the basis of rigorous kinetic
model of plasma one of the authors has shown that the frozen-in
condition is a consequence of {\it the drift approximation},
i.e., the smallness of Larmor length for electrons
$\lambda_e=c/\omega_c$ in comparison with the characteristic
inhomogeneity scale, $r$:
\begin{equation}\label{drift_ap}
\Lambda=\frac{c}{r\omega_e}\ll 1; \quad \omega_c=\frac{eH}{m_e},
\end{equation}
where $\omega_c$ is Larmor frequency for electrons.

\subsection{Self-consistent equations of magnetohydrodynamics}
The complete system of self-consistent RMHD equations for a
plasma in a gravitational field, obtained in \cite{Ign95},
consists of Maxwell equations of the first group:
\begin{equation}\label{1Maxwell}
\stackrel{*}{F}\ \!\!\!^{ik}_{~~,k}=0;
\end{equation}
Maxwell equations of the second group:
\begin{equation}\label{2Maxwell}
F^{ik}_{~~,k}=-4\pi J^i_{\rm{dr}}
\end{equation}
with spacelike {\it drift current}:
\begin{equation}\label{Jdr}
J^i_{\rm{dr}}=-\frac{2F^{ik} \stackrel{p}{T}\
\!\!\!^{l}_{k,l}}{F_{jm}F^{jm}};\quad (J_{\rm{dr}},J_{\rm{dr}})<0
\end{equation}
and a conservation law of the total energy-momentum of the system:
\begin{equation}\label{Tik,k}
\stackrel{p}{T}\ \!\!\!^{ik}_{~,k}+ \stackrel{f}{T}\
\!\!\!^{ik}_{~,k}=0.
\end{equation}
The continuity equation for the drift current must be satisfied
in consequence of Eq. (\ref{2Maxwell}):
\begin{equation}\label{cons_j}
J^i_{\rm{dr}~,i}=0.
\end{equation}
It should be noted some useful strict consequences of
magnetohydrodynamics equations:
\begin{eqnarray}
\label{F*J}\stackrel{*}{F}_{ik}J^k_{\rm{dr}}=0;\\
\label{vT}v^i  \stackrel{p}{T}\ \!\!\!^{ik}_{i,k}=0;\\
\label{HT}H^i\stackrel{p}{T}\ \!\!\!^{k}_{i,k}=0.
\end{eqnarray}

\subsection{Maxwell tensor representation by accompanying
intensities}
The components of Maxwell tensor is conveniently represented by a
pair of spacelike {\it vectors of accompanying intensities} of
electric, $E_i$ (\ref{embed}), and magnetic, $H_i$, fields
\cite{ZakIgn76}:
\begin{equation}\label{H}
E_i= F_{ji}v^j;\quad H_i=\stackrel{*}{F}_{ji}v^j,
\end{equation}
so that:
\begin{equation}\label{E^2,H^2}
(E,E)=-E^2; \quad (H,H)=-H^2;\quad (v,E)=0;\quad (v,H)=0.
\end{equation}
Then Maxwell tensor and dual to it can be expressed through a
pair of spacelike vectors of this accompanying intensities
\cite{ZakIgn76}:
\begin{eqnarray}\label{FEH}
F_{ij}=v_iE_j-v_jE_k-\eta_{ijkl}v^kH^l;\\
\label{F*EH} \stackrel{*}{F}_{ij}=v_iH_j-v_jH_k+\eta_{ijkl}v^kE^l,
\end{eqnarray}
where:
\begin{eqnarray}\label{FF}
\frac{1}{2}F_{ij}F^{ij}=\frac{1}{2}\stackrel{*}{F}_{ij} \stackrel{*}{F}\ \!\!\!^{ij}=(E,E)-(H,H)=H^2-E^2;\\
\frac{1}{2}F_{ij}\stackrel{*}{F}\ \!\!\!^{ij}=(E,H).
\end{eqnarray}
The energy-momentum tensor (EMT) of the electromagnetic field
\begin{equation}\label{T_f}
\stackrel{f}{T}\
\!\!\!^{i}_k=\frac{1}{4\pi}\left(F^i_{~l}F^l_{~k}+\frac{1}{4}\delta^i_k
F^{lm}F_{lm} \right) \end{equation}
can also be represented by the triplet of vectors $v, E, H$ (see
\cite{Ign95}). In the case of coincidence plasma's and
electromagnetic field's dynamic velocities (\ref{eq_vel}) the EMT
is expressed through a pair of vectors, $v, H$ \cite{Ign95}:
\begin{equation}\label{T_f_H}
\stackrel{f}{T}\ \!\!\!^{i}_k=-\frac{1}{8\pi}\left[(\delta^i_k-2v^i
v_k)H^2+2H^i H_k)\right],
\end{equation}
so:
\begin{equation}\label{SpTf}
\stackrel{f}{T}\equiv \stackrel{f}{T}\ \!\!\!^{i}_i=0.
\end{equation}
It is easy to verify that the vector $v$  and the spacelike unit
vector of the magnetic field $h$ --
\begin{equation}\label{h}
h^i=\frac{H^i}{H}; \quad (h,h)=-1;\quad (v,h)=0
\end{equation}
-- are actually the eigenvectors of the tensor $\stackrel{f}{T}\
\!\!\!^{ik}$:
\begin{eqnarray}\label{veH}
\stackrel{f}{T}\ \!\!\!^{i}_kv^k=\varepsilon_H v^i;\\
\label{heH} \stackrel{f}{T}\ \!\!\!^{i}_k h^k=\varepsilon_H h^i,
\end{eqnarray}
where the invariant
\begin{equation}\label{eH}
\varepsilon_H=\frac{H^2}{8\pi}
\end{equation}
is the energy density of the magnetic field.
\subsection{Energy-momentum tensor of magnetoactive plasma}
The energy-momentum tensor of a relativistic anisotropic
magnetoactive plasma in gravitational and magnetic fields is (see,
for example, \cite{IgnGor97}):
\begin{equation}\label{T_p}
\stackrel{p}{T}\ \!\!\!^{ij}=(\varepsilon +p_\perp)v^iv^j-p_\perp
g^{ij}+(p_\parallel -p_\perp)h^ih^j,
\end{equation}
where $p_\perp, p_\parallel$ - plasmas pressure in the directions
orthogonal and parallel to the magnetic field, respectively.
Trace of the energy-momentum tensor (\ref{T_p}) is:
\begin{equation}\label{Sp_T_p}
\stackrel{p}{T}\equiv\stackrel{p}{T}\
\!\!\!^{i}_i=\varepsilon-p_\perp-2p_\parallel\geq 0
\end{equation}
and because of the virial theorem (see \cite{land}) it is
non-negative:
\begin{equation}\label{ppe}
p_\perp+2p_\parallel\leq \varepsilon.
\end{equation}
It is easy to verify that the vectors $v$ and $h$ are also
eigenvectors of the energy-momentum tensor of plasma (see
(\ref{veH}), (\ref{heH}))
\begin{eqnarray}\label{T_p_v}
\stackrel{p}{T}\ \!\!\!^{ik} v_k=\varepsilon v^i;\\
\label{T_p_h} \stackrel{p}{T}\ \!\!\!^{ik} h_k=-p_\parallel h^i.
\end{eqnarray}

\section{Solving RMHD equations in the PGW metric}

\subsection{The metric of a plane gravitational wave\label{metric_s}}

The vacuum PGW metric is (see, for example, \cite{torn}):
\begin{equation}   \label{01}
d s^{2} = 2 du dv  - L^{2}d\Sigma^2,
\end{equation}
where:
\begin{equation}\label{dSigma}
d\Sigma^2=\cosh 2\gamma (e^{2 \beta}(dx^{2})^{2} + e^{-2\beta}(
dx^{3})^{2}) - 2 \sinh 2\gamma dx^2 dx^3
\end{equation}
- is a metric of ``plane''($x^2,x^3$); $\beta(u)$, $\gamma(u)$ -
amplitudes of the polarization $\mathbf{e}_+$ and
$\mathbf{e}_\times$, respectively; $u = \frac{1}{\sqrt{2}}(t -
x^{1})$ is the retarded time, $v= \frac{1}{\sqrt{2}}(t + x^{1})$ is
the advanced time. The amplitudes of PGW are arbitrary functions of
the retarded time $u$, and $L(u)$ is {\it a background factor} of
PGW, which defined by single nontrivial vacuum Einstein's
equation\footnote{The prime denotes differentiation with respect to
the retarded time $u$.}:
\begin{equation}\label{Eist_L}
L''+ L(\cosh^2 2\gamma\beta'^2+\gamma'^2)=0.
\end{equation}
At inversion of the coordinates in the plane ($x^2,x^3$) and
transformation of the PGW amplitude:
\begin{equation}\label{reflect}
x^2=x'^3;\quad x^3=x'^2;\quad \beta'=-\beta;\quad \gamma'=\gamma
\end{equation}
two-dimensional metric transforms into itself. Under rotations in
the plane ($x^2,x^3$) by the $\pi/4$ angle:
\begin{equation}\label{rot}
x^2=\frac{1}{\sqrt{2}}(x'^2+x'^3);\:
x^3=\frac{1}{\sqrt{2}}(x'^3+x'^2)
\end{equation}
two-dimensional metric is transformed to:
\begin{equation}
\begin{array}{l}
d\Sigma'^2=(\cosh 2\gamma\cosh 2\beta+\sinh
2\gamma)(dx'^2)^2+\\[5pt](\cosh 2\gamma\cosh
2\beta-\sinh 2\gamma)(dx'^3)^2+ \label{GW_45}2\cosh 2\gamma\sinh
2\beta dx'^2dx'^3.
\end{array}
\end{equation}
If $ \beta = 0 $, i.e., in the case of PGW with a single
polarization $\mathbf{e}_\times$, we get from (\ref{GW_45}):
$$d\Sigma'^2=e^{2\gamma}(dx'^2)^2+e^{-2\gamma}(dx'^3)^2$$
-- a PGW metric with a single polarization of $\mathbf{e}_+$.

For a weak gravitational wave:
\begin{equation}\label{weak_GW}
|\beta(u)|\ll 1;\; |\gamma(u)|\ll 1;\; L^2(u)=1+O^2(|\beta,\gamma|)
\end{equation}
rotation (\ref{rot}) is equivalent to the transformation of
inversion:
\begin{equation}\label{beta_gamma_rot}
\beta'=\gamma;\quad \gamma'=-\beta.
\end{equation}

\subsection{Initial conditions}
Let in the absence of PGW ($u\leq 0$):
\begin{equation}  \label{03}
\beta(u \leq 0)=0; \quad \beta'(u\leq 0)=0;\quad L(u \leq 0)=1,
\end{equation}
plasma is homogeneous and at rest:
\begin{eqnarray}\label{04a} v^v(u\leq 0)= v^u(u\leq 0)=
1/\sqrt{2};
\quad v^{2} =v^{3}=0;\nonumber\\
\label{04}\varepsilon(u \leq 0)=\stackrel{0}{\varepsilon}; \quad
p_\parallel(u \leq 0) = \stackrel{0}{p}_\parallel; \qquad p_\perp(u
\leq 0) = \stackrel{0}{p}_\perp,
\end{eqnarray}
and homogeneous magnetic field is directed in the $(x^{1},x^{2}) $
plane:
\begin{eqnarray}
\label{05a}H_1(u \leq 0)=\stackrel{0}{H} \cos\Omega\,; \quad H_2(u
\leq
0)=\stackrel{0}{H} \sin\Omega\,;\nonumber\\
\label{05} H_3(u \leq 0) = 0\,; \quad E_i(u \leq 0) = 0,
\end{eqnarray}
where $\Omega$ is the angle between the axis $0x^{1}$ (the PGW
propagation direction) and the magnetic field ${\bf H}$.

As we noted above, the effect of a PGW with polarization
$\mathbf{e}_+$ on a homogeneous plasma at the initial conditions
(\ref{04})-(\ref{05}) was considered in the quoted papers
\cite{Ign95}-\cite{IgnGor97}. Taking into account transformational
properties of the metric noted in section \ref{metric_s}, it means
that the effect of a monopolarized PGW on a homogeneous plasma has
been considered earlier when the projection of vector
$\stackrel{0}{\mathbf{H}}$ on the plane of PGW's front is parallel
to the polarization axis. The case, when this projection coincides
with the direction $x^2$ or $x^3$, can be reduced to the case of
polarization $\mathbf{e}_+$ or $\mathbf{e}_\times$ using the
substitution (\ref{reflect}). And vice versa: the case of different
polarizations $\mathbf{e}_+$ or $\mathbf{e}_\times$ can be reduced
to the case with different projections on the direction $x^2$ or
$x^3$ under rotation in the PGW's front plane by the angle $\pi/4$
together with rotation of the vector of magnetic field intensity.
For understanding of a mechanism of strong PGW interaction with an
anisotropic magnetoactive plasma it is essentially important to
consider the combined case, when a PGW possesses both polarization
states simultaneously, and the projection of the vector of magnetic
field intensity on the PGW's front plane is parallel to the axis of
one of them. The initial conditions (\ref{04})-(\ref{05}) correspond
to this case.

\subsection{Symmetry of the problem}

As is well known, the metric (\ref{01}) permits the group of
motions $G_{5}$, associated with three linearly independent in a
point Killing vectors:

\begin{equation} \label{02}
\mathop{\xi^{i}}\limits_{(1)} =\delta^{i}_{v}\,; \qquad
\mathop{\xi^{i}}\limits_{(2)} = \delta^{i}_{2}\,; \qquad
\mathop{\xi^{i}}\limits_{(3)} = \delta^{i}_{3}\,.
\end{equation}
In consequence of the Killing vectors existence in the metric
(\ref{01}), all of the geometric objects, including the
Christoffel symbols, the Riemann tensor, the Ricci tensor and,
consequently, the energy-momentum tensor of a magnetoactive
plasma, are automatically conserved at motions along the
Killing's directions:
\begin{equation}\label{symmetric}
\mathop{\mathrm{L}}\limits_{\xi_\alpha}g_{ij}=0;\Rightarrow
\mathop{\mathrm{L}}\limits_{\xi_\alpha}R_{ij}=0;\Rightarrow
\mathop{\mathrm{L}}\limits_{\xi_\alpha}T_{ij}=0,
\end{equation}
where $\mathop{\mathrm{L}}\limits_{\xi}T_{ij}$ is a Lie derivative
in the direction of $\xi$:
\begin{equation}\label{Lee}
\mathop{\mathrm{L}}\limits_{\xi}T_{ij}=T_{ij,k}
\xi^k+T_{kj}\xi_{,i}^{\ j}+T_{ik}\xi_{,j}^{\ k}.
\end{equation}
Further we require that tensors of energy-momentum of the plasma
$\stackrel{p}{T}_{ij}$ and the electromagnetic field
$\stackrel{f}{T}_{ij}$ inherit the symmetry separately:
\begin{eqnarray}\label{Lee_p}
\mathop{\mathrm{L}}\limits_{\xi_\alpha}\stackrel{p}{T}_{ij}=0;\\
\label{Lee_f}
\mathop{\mathrm{L}}\limits_{\xi_\alpha}\stackrel{f}{T}_{ij}=0; \quad
(\alpha=\overline{1,3}).
\end{eqnarray}
Consequences of (\ref{Lee_f}) are:
\begin{equation}\label{Lee_F}
\mathop{\mathrm{L}}\limits_{\xi_\alpha}F_{ij}=0,\;
\mathop{\mathrm{L}}\limits_{\xi_\alpha}\stackrel{*}{F}_{ij}=0\Longrightarrow
\mathop{\mathrm{L}}\limits_{\xi_\alpha}H=0,\:
\mathop{\mathrm{L}}\limits_{\xi_\alpha}E_i=0,\:
\mathop{\mathrm{L}}\limits_{\xi_\alpha}H_i=0.
\end{equation}
Consequences of (\ref{Lee_p}) and (\ref{Lee_F}) are:
\begin{equation}\label{Lee_P}
\mathop{\mathrm{L}}\limits_{\xi_\alpha}\varepsilon=0,\:
\mathop{\mathrm{L}}\limits_{\xi_\alpha}v^i=0,\:
\mathop{\mathrm{L}}\limits_{\xi_\alpha}p_\perp=0,\:\mathop{\mathrm{L}}\limits_{\xi_\alpha}p_\parallel=0.
\end{equation}
Thus, all observed physical quantities $\mathbf{P}$ {\em inherit the
symmetry of the metric} (\ref{01}):
\begin{equation}
\label{07} \mathop{\mathrm{L}}\limits_{\xi_\alpha} {\bf P} =0;
\hspace{1,5 cm} (\alpha =\overline{1,3}),
\end{equation}
i.e., taking into account the explicit form of Killing vectors
(\ref{02}):
\begin{eqnarray}
\label{08} p=p(u); \quad \varepsilon=\varepsilon(u); \quad v^{i}=v^{i}(u);\\
\label{09} F_{ik}=F_{ik}(u); \quad H_i=H_i(u); \quad h_i=h_i(u).
\end{eqnarray}

\subsection{Maxwell tensor}
In this section we obtain an expression for the vector potential
of the electromagnetic field in the metric (\ref{01}), taking
into account the initial conditions (\ref{04a})-(\ref{05}). This
method differs from the method used in \cite{Ign95}. It is based
only on the first group of Maxwell equations and the initial
conditions and therefore have greater generality. The vector
potential conformed with the initial conditions (\ref{05}) is:
\begin{eqnarray} A_{v} = A_{u} = A_{2} = 0;\nonumber\\
\label{06} A_{3}=\stackrel{0}{H} (x^{1} \sin\Omega - x^{2}
\cos\Omega); \qquad (u \leq 0).
\end{eqnarray}
These conditions conform with follow components of Maxwell
tensor:
\begin{eqnarray}\label{F_0}
F_{23}(u\leq 0)=-\stackrel{0}{H}\cos\Omega;\; F_{v3}(u\leq
0)=\frac{1}{\sqrt{2}}\stackrel{0}{H}\sin\Omega;\; F_{v2}(u\leq 0)=0;\nonumber\\
F_{u2}(u\leq 0)=0;\; F_{u3}(u\leq
0)=-\frac{1}{\sqrt{2}}\stackrel{0}{H}\sin\Omega;\; F_{uv}(u\leq
0)=0.
\end{eqnarray}
As known (see \cite{land}), the first group of Maxwell equations
(\ref{1Maxwell}) is equivalent to the existence condition of a
vector potential. It can be written as:
\begin{equation}\label{Maxwell_I}
\frac{1}{\sqrt{-g}}\partial_j \sqrt{-g}\stackrel{*}{F}\
\!\!\!^{ij}=0.
\end{equation}
Considering (\ref{09}), we get:
\begin{equation}
\label{10} L^{2} \stackrel{*}{F}\ \!\!\!^{u\alpha} = {\cal
C}_{(\alpha)} \quad (=\rm{Const}); \hspace{1 cm} \alpha = \lbrace
v,2,3 \rbrace,
\end{equation}
setting here and further the following order of the coordinates:
\begin{equation}\label{coord}
\rm{Coords}:=[v,u,x^2,x^3],
\end{equation}
Let us establish a connection between the components of Maxwell
tensor with the components of tensor dual to it:
\begin{eqnarray}\label{F*ik}
\stackrel{*}{F}\ \!\!\!^{uv}=-\frac{1}{L^2}F_{23};\quad
\stackrel{*}{F}\ \!\!\!^{u2}=\frac{1}{L^2}F_{v3};\quad \stackrel{*}{F}\ \!\!\!^{u3}=-\frac{1}{L^2}F_{v2};\nonumber\\
\stackrel{*}{F}\ \!\!\!^{v2}=\frac{1}{L^2}F_{u3};\quad
\stackrel{*}{F}\ \!\!\!^{v3}=\frac{1}{L^2}F_{u2};\quad
\stackrel{*}{F}\ \!\!\!^{23}=-\frac{1}{L^2}F_{uv}.
\end{eqnarray}
Then the initial conditions (\ref{05}) give:
\begin{eqnarray}\label{11a}
L^{2} \stackrel{*}{F}\ \!\!\!^{uv}= - F_{23}= \stackrel{0}{H} \cos\Omega ;\\
\label{11b}L^{2}\stackrel{*}{F}\ \!\!\!^{u2}= F_{v3}= \frac{1}{\sqrt2} \stackrel{0}{H}\sin\Omega ;\\
\label{11} L^{2} \stackrel{*}{F}\ \!\!\!^{u3}= -F_{v2}= 0.
\end{eqnarray}
Thus, the second invariant of the electromagnetic field is equal to:
\begin{equation}\label{2inv}
{\rm Inv}_2=F_{ik}\stackrel{*}{F}\
\!\!\!^{ik}=\frac{2}{L^2}(F_{v3}F_{u2}-F_{23}F_{uv}),
\end{equation}
So, taking into account (\ref{11b}), (\ref{11}), the equality to
zero of the second invariant of an electromagnetic field
(\ref{II_inv}) is reduced to the relation:
\begin{equation}
\label{12} L^{2} \stackrel{*}{F}\ \!\!\!^{v3}\equiv F_{u2}=
-{\sqrt 2} F_{uv} \cot\Omega.
\end{equation}
As it is known (see for example \cite{land}), the first group of
Maxwell equations is equivalent to the existence condition of a
vector potential $A_{i}$:
\begin{equation}
\label{13} F_{ik} = \partial_{i} A_{k} - \partial_{k} A_{i}.
\end{equation}
Let us notice that as opposed to Maxwell tensor, the components
of the vector potential $A_i$ can depend on the variables $v,
x^2, x^3$. We write down expressions (\ref{11a})-(\ref{12})
relative to the vector potential $A_i$ using definition of
Maxwell tensor (\ref{13}):
\begin{eqnarray}
\label{A23}\partial_3A_2-\partial_2A_3=\stackrel{0}{H}\cos\Omega;\\
\label{Av3}\partial_vA_3-\partial_3A_v=\frac{1}{\sqrt{2}}\stackrel{0}{H}\sin\Omega;\\
\label{Av2}\partial_vA_2-\partial_2A_v=0;\\
\label{Au2v}\partial_uA_2-\partial_2A_u=-\sqrt{2}\cot\Omega(\partial_uA_v-\partial_vA_u))
\end{eqnarray}
Introducing new functions:
\begin{eqnarray}
\label{tildeA2}\tilde{A}_2=A_2-\stackrel{0}{H}\cos\Omega \ x^3
\equiv A_2-\delta
A_2;\\
\label{tildeAv}
\tilde{A}_v=A_v+\frac{1}{\sqrt{2}}\stackrel{0}{H}\sin\Omega\
x^3\equiv A_v-\delta A_v,\\
\label{tildeA3}\tilde{A}_3=A_3,
\end{eqnarray}
where:
\begin{equation}\label{dA}
\delta A_2=\stackrel{0}{H}\cos\Omega \ x^3;\quad \delta
A_v=-\frac{1}{\sqrt{2}}\stackrel{0}{H}\sin\Omega\ x^3;\quad \delta
A_3=0,
\end{equation}
let us reduce the relations (\ref{A23}) and (\ref{Av3}) to the form
similar to (\ref{Av2}):
\begin{eqnarray}
\label{tildeA23}\partial_3 \tilde{A}_2-\partial_2 A_3=0;\\
\label{tildeAv3} \partial_v A_3-\partial_3 \tilde{A}_v=0.
\end{eqnarray}
Let us notice that the renormalization of the component of the
vector potential (\ref{tildeA2}), (\ref{tildeAv}) keeps the
relation (\ref{Av2}) invariable. But then it is possible to write
down the system of equations (\ref{Av2}), (\ref{tildeA23}),
(\ref{tildeAv3}) as:
\begin{equation}\label{tildeAv23}
\partial_\sigma \tilde{A}_\delta-\partial_\delta \tilde{A}_\sigma=0;\quad (\sigma,\delta=v,2,3)
\end{equation}
and to consider it as equations on a three-dimensional hypersurface
$V^3=\{v,x^2,x^3\}$. As it is known, the unique solution of
equations (\ref{tildeAv23}) on $V^3$ is a gradient function:
\begin{equation}
\label{14} \tilde{A}_{\sigma} = \partial_{\sigma}\Phi, \quad
(\sigma=v,2,3),
\end{equation}
where $\Phi=\Phi(u,v,x^2,x^3)$ is an arbitrary scalar function. The
value of the potential function corresponding to the initial
conditions (\ref{06}) is:
\begin{equation}\label{Phi0}
\Phi(u\leq
0)=x^3\stackrel{0}{H}\left(\frac{1}{\sqrt{2}}(v-u)\sin\Omega
-x^2\cos\Omega\right).
\end{equation}
Thus
\begin{equation}\label{AF}
A_\sigma=\partial_\sigma \Phi+\delta A_\sigma.
\end{equation}

As it is known (see, for example, \cite{land}), it is possible to
impose one gauge condition on 4 components of a vector potential. We
choose this condition in the form corresponding to the initial
conditions (\ref{06}):
\begin{equation}\label{Au}
A_u=0.
\end{equation}
Then for the nonconserved components of the Maxwell tensor
$F_{u\sigma}$ is valid:
\begin{equation}\label{Fus}
F_{u\sigma}=\partial_{u\sigma}\Phi; \quad (\sigma=v,2,3).
\end{equation}
But then condition (\ref{Au2v}) can be written down in the form:
\begin{equation}\label{A2Av}
\partial_u(A_2+\sqrt{2}\cot\Omega A_v)=0.
\end{equation}
Integrating (\ref{A2Av}) with the initial conditions (\ref{06}), we
obtain:
\begin{equation}\label{A2+Av}
A_2+\sqrt{2}\cot\Omega A_v=0.
\end{equation}
Taking into account the identity:
\begin{equation}\delta A_2+\sqrt{2}\cot\Omega\delta A_v\equiv0,\end{equation}
we obtain the linear differential equation from (\ref{A2+Av}):
$$\partial_2\Phi+\sqrt{2}\cot\Omega
\partial_v\Phi=0.$$
Integrating it, we obtain:
\begin{equation}\label{Phi_int}
\Phi=\Phi(v\sqrt{2}\sin\Omega-x^2\cos\Omega,u,x^3),
\end{equation}
where $\Phi$ is an arbitrary function of its arguments. Using now
the initial condition (\ref{Phi0}) we obtain finally:
\begin{equation}\label{Phi}
\Phi=x^3\stackrel{0}{H}\left(\frac{1}{\sqrt{2}}(v-\psi(u))\sin\Omega
-x^2\cos\Omega\right),
\end{equation}
where $\psi(u)$ is an arbitrary function of the retarded time,
satisfying the initial condition:
\begin{equation}\label{phi0}
\psi(u\leq 0)=u.
\end{equation}
Thus, the final expression for components of the vector potential
becomes:
\begin{equation}\label{Ai}
A_2=A_v=A_u=0;\quad
A_3=\stackrel{0}{H}\left(\frac{1}{\sqrt{2}}(v-\psi(u))\sin\Omega-x^2\cos\Omega
\right).
\end{equation}
The components of the Maxwell tensor relative to the potential
(\ref{Ai}) are equal to:
\begin{eqnarray}\label{Fik}
F_{vu}=0;\quad F_{2u}=0;\quad
F_{3u}=\frac{1}{\sqrt{2}}\stackrel{0}{H}\psi'\sin\Omega;\nonumber\\
F_{2v}=0;\quad
F_{3v}=-\frac{1}{\sqrt{2}}\stackrel{0}{H}\sin\Omega;\quad
F_{23}=-\stackrel{0}{H}\cos\Omega
\end{eqnarray}
and are defined only by one unknown function $\psi(u)$. For the
components of the dual Maxwell tensor (\ref{F_dual}) we get:
\begin{eqnarray}\label{F*ik}
\stackrel{*}{F}\ \!\!\!^{vu} =
\frac{1}{L^2}\stackrel{0}{H}\cos\Omega;\quad \stackrel{*}{F}\
\!\!\!^{2u} = -\frac{1}{\sqrt{2}L^2}\stackrel{0}{H}\sin\Omega;\quad
\stackrel{*}{F}\ \!\!\!^{3u} = 0;\nonumber\\
\stackrel{*}{F}\ \!\!\!^{2v} =
\frac{1}{\sqrt{2}L^2}\stackrel{0}{H}\psi'\sin\Omega;\quad
\stackrel{*}{F}\ \!\!\!^{3v}=0;\quad \stackrel{*}{F}\ \!\!\!^{23}=0.
\end{eqnarray}

\subsection{Accompanying intensities and the frozen-in condition}
According to (\ref{H}), we define the components of the vector of
accompanying intensity of the electric field, $E_i$, as:
\begin{eqnarray}\label{Ei}
E_v=-\frac{1}{\sqrt{2}}\stackrel{0}{H}\sin\Omega\ v^3; \quad %
E_u=\frac{1}{\sqrt{2}}\stackrel{0}{H}\psi'\sin\Omega\ v^3;\nonumber \\
E_2=\stackrel{0}{H}\cos\Omega\ v^3;\quad
E_3=\frac{1}{\sqrt{2}}\stackrel{0}{H}\sin\Omega (v^v-\psi'
v^u)-\stackrel{0}{H}\cos\Omega\ v^2.
\end{eqnarray}
Thus, the frozen-in condition of magnetic field in plasma
(\ref{embed}) reduces to two equalities:
\begin{equation}\label{vi}
v^3=0;\quad
\frac{1}{\sqrt{2}}(v_v\psi'-v_u)\sin\Omega+v^2\cos\Omega=0.
\end{equation}
As a result, covariant components of the Maxwell tensor,
contravariant components of the dual Maxwell tensor and
contravariant components of the velocity vector are defined by
the expressions, obtained in \cite{Ign95}, but now they are
already defined for a more general metric of a gravitational wave
and at weaker assumptions. In the quoted paper, in particular, to
obtain the explicit form of the Maxwell tensor components and the
velocity vector components, the analysis of the drift current
components was carried out using the conservation law of this
current. As it was shown above, for achievement of this purpose
three assumptions are
sufficient:\\
1. inheritance of the space symmetry by the energy-momentum
tensor of the electromagnetic field and by the
energy-momentum tensor of the plasma separately;\\
2. the equality to zero of the second invariant of the Maxwell tensor;\\
3. the frozen-in condition of magnetic field in plasma.\\

Thus, the analysis of the first group of Maxwell equations and
initial conditions is sufficient.

Calculating further the covariant components of the dual Maxwell
tensor, subject to (\ref{F*ik}), we get:
\begin{eqnarray}\label{F*_ik}
\hspace{-1cm}
\stackrel{\ast}{F}_{uv}=\frac{\stackrel{0}{H}}{L^2}\cos\Omega;\;
\stackrel{\ast}{F}_{u2}=\frac{\stackrel{0}{H}}{\sqrt{2}}e^{2\beta}\cosh2\gamma
\psi'\sin\Omega;\:\stackrel{\ast}{F}_{v2}=-\frac{\stackrel{0}{H}}{\sqrt{2}}
e^{2\beta}\cosh2\gamma\sin\Omega;\nonumber\\
\hspace{-1cm}
\stackrel{\ast}{F}_{v3}=\frac{\stackrel{0}{H}}{\sqrt{2}}
\sinh2\gamma\sin\Omega;\quad
\stackrel{\ast}{F}_{u3}=-\frac{\stackrel{0}{H}}{\sqrt{2}}\sinh2\gamma
\psi'\sin\Omega;\quad\stackrel{\ast}{F}_{23}=0.
\end{eqnarray}

Covariant components of the vector of magnetic field intensity
relative to the Maxwell tensor (\ref{F*ik}) are equal to:
\begin{eqnarray}\label{Hv}
H_v=-\frac{\stackrel{0}{H}}{L^2} \left(v_v
\cos\Omega+\frac{1}{\sqrt{2}} v^2 \sin\Omega \right);\\
\label{Hu} H_u = \frac{\stackrel{0}{H}}{L^2} \left( v_u \cos\Omega
- \frac{1}{\sqrt{2}}v^2 \psi' \sin\Omega \right);\\
\label{H2} H_2 = -\frac{1}{\sqrt{2}} \stackrel{0}{H} \cosh2\gamma e^{2\beta} \sin\Omega (v_v \psi'+ v_u );\\
\label{H3} H_3 = \frac{1}{\sqrt{2}} \stackrel{0}{H} \sinh2\gamma
\sin\Omega (v_v \psi'+ v_u ).
\end{eqnarray}
It is thus easy to show on the basis of formula (\ref{F*ik}):
\begin{equation}\label{H^3}
H^3=\stackrel{*}{F}\ \!\!\!^{i3}v_i=0,
\end{equation}
i.e., the third contravariant coordinate of the vector of the
magnetic field intensity, as well as the vector of dynamic
velocity of the plasma, is equal to zero. Also it is easy to be
convinced of orthogonality of the velocity vector and the
magnetic field intensity (\ref{E^2,H^2}):
\begin{equation}\label{vH}
H_iv^i\equiv0.
\end{equation}
The square of the magnetic field intensity, i.e., a scalar $H^2$,
most easier to calculate by means of the relation (\ref{FF}),
using the explicit form of contravariant (\ref{F*ik}) and
covariant (\ref{F*_ik}) components of the dual Maxwell tensor:
\begin{equation} \label{33}
H^2 = \frac{\stackrel{0}{H}\ \!\!\!^{2}}{L^4} (L^2 \psi'
\cosh2\gamma e^{2\beta} \sin^2\Omega + \cos^2\Omega )\,.
\end{equation}
The frozen-in conditions of magnetic field in plasma (\ref{vi})
establishes the connection between nonzero contravariant components
of the velocity vector $v^2, v^v=v_u, v^u=v_v$. Besides, there is
still the normalization relation of velocity vector (\ref{eq_vel}).
Therefore, the only one independent coordinate of the velocity
vector remains, and the electromagnetic field is defined by only one
unknown function of the retarded time, $\psi(u)$. Using
(\ref{Hv})-(\ref{33}) the normalization relation of velocity vector
can be written in the equivalent form:
\begin{equation}  \label{34}
\left[ v_v \cos\Omega + v_2 \frac{1}{\sqrt{2}} \sin\Omega
\right]^2 = \frac{H^2}{\stackrel{0}{H}\ \!\!\!^{2}} v^2_v L^4 -
\frac{\sin^2 \Omega}{2} L^2 \cosh2\gamma e^{2\beta}\,.
\end{equation}

\subsection{Drift current}
Let us calculate components of a drift current, using Maxwell
equations (\ref{2Maxwell}), considering  the dependence of Maxwell
tensor components only on the retarded time (\ref{09}):
\begin{equation}\label{curr}
J^i_{\rm{dr}} = -\frac{1}{4\pi L^2}\partial_u (L^2 F^{iu}).
\end{equation}
Then:
\begin{equation}\label{J^v}
J^v_{\rm{dr}} = J^u_{\rm{dr}} = 0\,;
\end{equation}
\begin{equation}\label{J^2}
J^2_{\rm{dr}} = -\frac{\stackrel{0}{H}\sin\Omega}{2\sqrt{2}\pi L^2}
\cosh2\gamma\cdot\gamma' \,;
\end{equation}
\begin{equation}\label{J^3}
J^3_{\rm{dr}} = -\frac{\stackrel{0}{H}\sin\Omega
e^{2\beta}}{2\sqrt{2}\pi L^2}
(\sinh2\gamma\cdot\gamma'+\cosh2\gamma\cdot\beta') \,.
\end{equation}
Calculating scalar product of the vector of the magnetic field
intensity and the vector of drift current density, using
(\ref{H2}), (\ref{H3}), (\ref{J^v}), (\ref{J^2}), (\ref{J^3}), we
get:
\begin{equation}\label{JH}
(J_{\rm{dr}},H)=\frac{\stackrel{0}{H}\ \!\!\!^{2}}{4\pi L^2}(v_v
\psi'+ v_u )(\gamma'-\frac{\beta'}{2}\sinh4\gamma).
\end{equation}
Thus, the presence of the second polarization of a gravitational
wave leads to violation of the orthogonality of the vectors of drift
current density and magnetic field intensity\footnote{We remind that
in case of monopolarized gravitational waves these spacelike vectors
are mutually orthogonal \cite{Ign95}.}.

Using expression (\ref{Jdr}), it is possible to show that the
equality ({\ref{J^v}}) is carried out only in the case of transverse
propagation of the PGW ($\Omega=\pi/2$).

\subsection{Integrals of the motion}
Because of existence of the motions (\ref{02}), Killing equations
are satisfied:
\begin{equation}\label{Killing}
\mathop{\xi}\limits_{(\alpha)}{}_{i,k}+\mathop{\xi}\limits_{(\alpha)}{}_{k,i}=0,\quad
(\alpha=\overline{1,3}).
\end{equation}
Therefore conservation laws of the total EMT in a field of PGW after
consistently transvection with all Killing's vectors (\ref{02}) can
be written down in the form:
\begin{equation}
\frac{1}{\sqrt{-g}}(\partial_k
\sqrt{-g}{\mathop{\xi}\limits_{(\alpha)}}^i T^k_i)=0; \quad
(\alpha=\overline{1,3}).
\end{equation}
Taking into account the fact that EMT components can depend only on
the retarded time, we obtain following integrals \cite{Ign95}:
\begin{equation}  \label{35}
L^2 \mathop{\xi}\limits_{(\alpha)}{}^i T_{v i} = C_a = {\rm Const};
\quad (\alpha = \overline{1,3})\,.
\end{equation}
In this paper we consider only the case of {\it transverse
propagation} of the PGW ($\Omega=\pi/2$). Then, substituting
expressions for the EMT of the plasma and electromagnetic field in
the integrals (\ref{35}), using relations (\ref{H3})-(\ref{34}) and
also initial conditions (\ref{03}), (\ref{04}), we lead integrals of
the motion to the form:
\begin{equation}  \label{C1}
2 L^2 (\varepsilon + p_\parallel) v_v^2 - (p_\parallel - p_\perp)
\frac{\stackrel{0}{H}\ \!\!\!^{2}}{H^2} \cosh2\gamma e^{2\beta} =
(\stackrel{0}{\varepsilon} + \stackrel{0}{p}) \Delta(u)\,;
\end{equation}
\begin{equation}  \label{C2}
L^2 (\varepsilon + p_\parallel) v_v v_2 = 0 \,;
\end{equation}
\begin{equation}  \label{C3}
L^2 (\varepsilon + p_\parallel) v_v v_3 = 0 \,,
\end{equation}
where:
\begin{equation}  \label{38}
\stackrel{0}{p} = \stackrel{0}{p}_\perp\,;
\end{equation}
and so-called {\it the governing function of GMSW} is introduced:
\begin{equation}  \label{40}
\Delta(u) =   1 - \alpha^2 (\cosh2\gamma e^{2\beta} - 1)\,,
\end{equation}
with {\it dimensionless parameter} $\alpha^2$:
\begin{equation}  \label{alpha}
\alpha^2 = \frac{\stackrel{0}{H}\ \!\!\!^{2}}{4\pi
(\stackrel{0}{\varepsilon} + \stackrel{0}{p})}\,.
\end{equation}
Solving (\ref{C1}) with respect to $v_v$ we obtain expressions for
coordinates of the velocity vector as functions of the scalars :
$\varepsilon$, $p_\parallel$, $p_\perp$, $\psi'$ and the explicit
functions of the retarded time:
\begin{equation}  \label{Vv}
v_v^2 =  \frac{(\stackrel{0}{\varepsilon} +
\stackrel{0}{p})}{2L^2(\varepsilon + p_\parallel)} \Delta(u) +
\frac{(p_\parallel - p_\perp)}{(\varepsilon + p_\parallel)}
\frac{\stackrel{0}{H}\ \!\!\!^{2}}{H^2} \frac{\cosh2\gamma
e^{2\beta}}{2L^2}\,;
\end{equation}
From (\ref{C1}) we get:
\begin{equation}  \label{V2}
v_2 = 0 \,.
\end{equation}
We obtain the coordinate $v_u$ from a normalization relation of
velocity vector, using (\ref{Vv}), (\ref{V2}) :
\begin{equation}  \label{Vu}
v_u = \frac{1}{2 v_v}\,,
\end{equation}
and from the frozen-in condition (\ref{vi}) we get the value of a
derivative of potential $\psi'$:
\begin{equation}  \label{psi}
\psi' = \frac{1}{2 v_v^2}\,,
\end{equation}
using it, the scalar $H^2$ is defined from relation (\ref{33})
as:
\begin{equation}  \label{H^2(perp)}
H^2 = \frac{\stackrel{0}{H}\ \!\!\!^{2}}{L^2} \frac{\cosh2\gamma
e^{2\beta}}{2 v_v^2}.
\end{equation}
Let us notice that in the case of an isotropic plasma
($p_{\perp}=p_{\parallel}=p$) the expression (\ref{Vv}) becomes:
\begin{equation}
v_v^2 =  \frac{(\stackrel{0}{\varepsilon} +
\stackrel{0}{p})}{2L^2(\varepsilon + p)} \Delta(u)\,;
\end{equation}
From RMHD system of equations it is possible to obtain a following
differential equation in the PGW metric:
\begin{equation}  \label{47}
L^2 \varepsilon' v_v + (\varepsilon + p_\parallel)(L^2 v_v)' +
\frac{1}{2}L^2 (p_\parallel - p_\perp) v_v (\ln H^2)' = 0\,.
\end{equation}
Finally, the equation (\ref{47}) is the differential equation on
3 unknown scalar functions: $\varepsilon$, $p_\parallel$ and
$p_\perp$. Such underdefiniteness is a known consequence of the
incompleteness of hydrodynamic description of a plasma. To solve
this equation it is necessary to impose two additional
connections between functions $\varepsilon$, $p_\parallel$,
$p_\perp$, i.e., an equation of state:
\begin{equation}  \label{48}
p_\parallel = f(\varepsilon)\,; \quad p_\perp = g(\varepsilon)\,.
\end{equation}

\section{Barotropic equation of state}

\subsection{General formulas}
Let us consider a barotropic state of an anisotropic plasma, when
the connections (\ref{48}) are linear:
\begin{equation}    \label{49}
p_\parallel = k_\parallel \varepsilon \,; \quad p_\perp = k_\perp
\varepsilon\,,
\end{equation}
The equation (\ref{47}) is easy to integrate at the connections
(\ref{49}), and we get one more integral:
\begin{equation}  \label{50}
\varepsilon (\sqrt{2} L^2 v_v)^{(1 + k_\parallel)} H^{(k_\parallel -
k_\perp)} = \stackrel{0}{\varepsilon} \stackrel{0}{H}\
\!\!\!^{(k_\parallel - k_\perp)}\,.
\end{equation}
Thus, formally the problem is solved, as it is reduced to the
solution of the algebraic equations system which, however, is
still too difficult to solve and analyse. The solution is
essentially defined by two dimensionless parameters: $k_\perp$
and $k_\parallel$. Further we consider the special cases of these
parameters.

\subsection{Transverse propagation of the PGW}
In the case of a barotropic equation of state at the connections
(\ref{49}) substitution of (\ref{H^2(perp)}) in (\ref{Vv}) leads
to result:
\begin{equation}  \label{54}
v^2_v = \frac{1}{2}\frac{\stackrel{0}{\varepsilon}}{L^2 \varepsilon}
\Delta (u)\,.
\end{equation}
Substituting (\ref{H^2(perp)}), (\ref{54}) in (\ref{50}), we
obtain the closed equation relative to the variable
$\varepsilon$, solving which, we get definitively:
\begin{equation}  \label{bar_E}
\varepsilon = \stackrel{0}{\varepsilon} \Big[ \Delta^{1+k_\perp}
L^{2(1+k_\parallel)} (\cosh2\gamma
e^{2\beta})^{k_\parallel-k_\perp} \Big]^{-g_\perp} \,;
\end{equation}
\begin{equation}  \label{bar_Vv}
v_v = \frac{1}{\sqrt{2}} \left[ \Delta L^{(k_\parallel+k_\perp)}
(\cosh2\gamma e^{2\beta})^{\frac{k_\parallel-k_\perp}{2}}
\right]^{g_\perp}\,;
\end{equation}
\begin{equation}  \label{bar_H} \displaystyle
H = \stackrel{0}{H} \left[ \Delta L^{(1+k_\parallel)}
(\cosh2\gamma e^{2\beta})^{-\frac{1-k_\parallel}{2}}
\right]^{-g_\perp}\,,
\end{equation}
where
\begin{equation}  \label{58}
g_\perp = \frac{1}{1 - k_\perp} \in [1, 2]\,.
\end{equation}
In particular, for ultrarelativistic plasma with zero parallel
pressure:
\begin{equation}  \label{59}
k_\parallel \rightarrow 0\,; \quad k_\perp \rightarrow \frac{1}{2}
\end{equation}
we obtain from (\ref{bar_E})-(\ref{58}):
\begin{equation}  \label{60}
v_v = \frac{1}{\sqrt{2}} L \Delta^2 (\cosh2\gamma
e^{2\beta})^{-1/2}\,;
\end{equation}
\begin{equation}  \label{61}
\varepsilon = \stackrel{0}{\varepsilon} L^{-4} \Delta^{-3}
(\cosh2\gamma e^{2\beta})\,; \quad H = \stackrel{0}{H} L^{-2}
\Delta^{-2} (\cosh2\gamma e^{2\beta})\,.
\end{equation}

\section{Energy balance equation}
In \cite{Ign95} was shown that the singular state, which exists in a
magnetized plasma under the condition $2 \beta_0 \alpha^2 > 1$ on
the hypersurface:
\begin{equation}  \label{62}
\Delta(u_*) = 0\,,
\end{equation}
is removed using the back effect of the magnetoactive plasma on the
GW. That leads to the efficient absorption of GW energy by the
plasma and restriction on the amplitude of the GW. An exact solution
of the PGW energy transformation to the energy of the shock wave is
possible only on the basis of the self-consistent system of
Einstein's equations and magnetohydrodynamics equations. However,
qualitative analysis of this situation can be carried out using a
simple model of energy balance proposed in \cite{Ign96}. The energy
flow of the magnetoactive plasma is directed along the direction of
the PGW propagation, i.e., along the axis $0x^{1}$. Let $\beta_*(u)$
and $\gamma_*(u)$ are the vacuum amplitudes of the PGW. In
WKB-approximation:
\begin{equation}\label{WKB}
8\pi\varepsilon \ll \omega^2\,,
\end{equation}
where $\omega$ is the characteristic PGW frequency and $\varepsilon$
is the matter energy density, all the functions still depend only on
the retarded time (see \cite{IgnBal81}). Thus, $\beta(u)$ and
$\gamma(u)$ are the amplitudes of the PGW subject to absorption in
plasmas; $T^{ij}$ is the total energy-momentum tensor of the plasma
and the electromagnetic field (\ref{TEM}).

\subsection{Integral law of energy conservation}
Ref. \cite{Ign95} suggested a semiquantitative  solution of this
problem on the basis of a simple model of energy balance. Due to its
extreme importance, we do not restrict ourselves to \cite{Ign95} and
return to a more complete study of the problem of energy
transmission from a GW to magnetoactive plasma. However, instead of
solving Einstein's equations, we make use of their consequence, the
conservation law of the total momentum of the system ``plasma +
gravitational waves''. Clearly, this model is only approximate and
cannot replace a rigorous solution of Einstein's equations.
According to \cite{land}, an arbitrary gravitational field provides
the conservation of the system's momentum:
\begin{equation}  \label{GSMW602}
p^i = \frac{1}{c}\int(-g)(T^{i4} + \stackrel{g}{T}{}^{i4})dV,
\end{equation}
where $\stackrel{g}{T}{}^{ik}$ is the energy-momentum pseudotensor
of the gravitational field and the integration covers the whole
3-dimensional space. Let us take into account that the above
solution is plane-symmetric and depends on the retarded time $u$
only. Consequently the integration over the ``plane'' $(x^2,x^3)$ in
(\ref{GSMW602}) reduce to simply multiplying by an infinite
2-dimensional area. Dividing both sides of (\ref{GSMW602}) by this
area and bearing in mind that with $\Omega = \pi/2$ among the
3-dimensional flow only $P^1$is nonzero, we obtain the conservation
law of the surface density of the momentum $P^1_{\Sigma}$:
\begin{equation}  \label{GSMW603}
P^1_{\Sigma} = \frac{1}{c}\int\limits_{-\infty}^{+\infty}(-g)(T^{14}
+ \stackrel{g}{T}{}^{14})dx \quad (= {\rm Const}).
\end{equation}
Let the right semispace $x > 0$ be filled with magnetoactive plasma
and the left one $x < 0$ with matter which does not interact with a
weak GW. Let further the whole gravitational momentum be
concentrated in the interval $u\in [0,u_f]$ where $t_f =
\sqrt{2}u_f$ is the gravitational pulse duration. Since the integral
in Eq. (\ref{GSMW603}) is conserved all the time, let us consider it
at $t_0 < 0$, when the GW has not yet reached the magnetoactive
plasma, and $-t_f > t> 0)$, when the GW has reached the plasma.
Taking into account that the vacuum solution depends only on the
retarded time, we get for the integral in Eq. (\ref{GSMW603}):
\begin{equation}  \label{GSMW604}
\int\limits_{0}^{u_f}\stackrel{g}{T}{}^{14}_0du =
\int\limits_{0}^{t\sqrt{2}}(T^{14} + \stackrel{g}{T}{}^{14})du +
\int\limits_{t/\sqrt{2}}^{u_f}\stackrel{g}{T}{}^{14}_0du,
\end{equation}
where $\stackrel{g}{T}{}^{14}_0 = \stackrel{g}{T}{}^{14}(\beta_*(u),
\gamma_*(u))$; $\stackrel{g}{T}{}^{14} =
\stackrel{g}{T}{}^{14}(\beta(u), \gamma(u))$. Transferring one of
the integrals to the left-hand side of Eq.(\ref{GSMW604}), we obtain
the relation:
\begin{equation}  \label{GSMW605}
\int\limits_{0}^{u}\stackrel{g}{T}{}^{14}_0du =
\int\limits_{0}^{u}(T^{14} + \stackrel{g}{T}{}^{14})du,
\end{equation}
where the variable $u = t/\sqrt{2} > 0$ can now take {\it any
positive} values.

A similar law may be written for the plasma total energy; in this
case instead of Eq.(\ref{GSMW605}) we obtain:
$$\int\limits_{0}^{u}\stackrel{g}{T}{}^{44}_0du = \int\limits_{0}^{u}(T^{44} -{\cal E}_0 + \stackrel{g}{T}{}^{14})du,$$
where ${\cal E}_0$ is the total energy density of the unperturbed
plasma.

\subsection{Local analysis of the conservation law}

Since the relation (\ref{GSMW605}) must be valid at any values of
the variable $u$, the corresponding local relation should be
satisfied:
\begin{equation}\label{eq_T_41}
T^{41}(\beta,\gamma) + \stackrel{g}{T}{}^{41}(\beta,\gamma) =
\stackrel{g}{T}{}^{41}(\beta_*,\gamma_*)\, ,
\end{equation}
where $\stackrel{g}{T}{}^{41}(\beta,\gamma)$ is the energy flow of a
weak GW in the direction $0x^{1}$ (see \cite{land}):
\begin{equation}\label{T_GW_41}
\stackrel{g}{T}{}^{41} = \frac{1}{16 \pi} \Big[ h'^{2}_{23} +
\frac{1}{4} (h'_{22} - h'_{33})^2 \Big] = \frac{1}{4\pi}
\big[(\gamma')^2 + (\beta')^2\big].
\end{equation}
The prime denotes differentiation with respect to $s$. At
substituting (\ref{T_GW_41}) into (\ref{eq_T_41}) and changing
variables to $v, u$, we obtain:
\begin{equation}\label{eq_T_temp}
2\pi \Big[ T^{vv} - T^{uu} \Big] + (\gamma')^2 + (\beta')^2 =
(\gamma'_*)^2 + (\beta'_*)^2.
\end{equation}
In the case of transversal PGW propagation and at a barotropic
equation of state of an anisotropic plasma we obtain:
\begin{equation}
T^{vv} - T^{uu} = \left( \frac{1}{4 v_v^2 } - v_v^2 \right) \left(
\varepsilon (1 + k_{\perp} ) + \frac{H^2}{4 \pi} \right).
\end{equation}
Further, using the solutions of magnetohydrodynamics for a
barotropic equation of state of plasma (\ref{bar_E}),
(\ref{bar_Vv}), (\ref{bar_H}) and dimensionless parameter
$\alpha^2$ (\ref{alpha}), we rewrite the energy balance equation
(\ref{eq_T_temp}) as:
$$
\frac{\stackrel{0}{H}\ \!\!\!^{2}}{4 L^2} \left(
\Delta^{-\frac{4}{1-k_\perp}} L^{-\frac{4
(k_\parallel+k_\perp)}{1-k_\perp}} (\cosh2\gamma
e^{2\beta})^{-\frac{2 (k_\parallel-k_\perp)}{1-k_\perp}} -
1\right) \left(\frac{1}{\alpha^2} + 1\right)+
$$
\begin{equation}\label{eq_T_temp1}
(\gamma')^2 + (\beta')^2 = (\gamma'_*)^2 + (\beta'_*)^2.
\end{equation}
Let us expand the expression in brackets by the smallness of the PGW
amplitudes (\ref{weak_GW}) but hold the term with $\Delta^{-1}$,
since the parameter $\alpha^2$ in a strongly magnetized plasmas can
be so large that the condition $2\alpha^2\beta > 1$ is satisfied.
Then energy balance equation takes the form:
\begin{equation}\label{eq_T_temp2}
\frac{\stackrel{0}{H}\ \!\!\!^{2}}{4}\left( \Delta^{-4 g_\perp} -
1\right) \left(\frac{1}{\alpha^2} + 1\right) + (\gamma')^2 +
(\beta')^2 = (\gamma'_*)^2 + (\beta'_*)^2.
\end{equation}
Since in linear approximation by the smallness of the amplitudes
$\beta$ and $\gamma$ the governing function (\ref{40}) does not
depend on the function $\gamma(u)$:
\begin{equation}  \label{65}
\Delta(u) = 1 - 2 \alpha^2 \beta  +O(\beta^2,\gamma^2)\,,
\end{equation}
and the functions $\beta(u)$, $\gamma(u)$ are arbitrary and
functionally independent, then up to $\beta^2, \gamma^2$, the
relation (\ref{eq_T_temp2}) should be decompose into two independent
parts:
\begin{eqnarray}\label{b}
2\stackrel{0}{H}\ \!\!\!^{2} g_\perp (1+\alpha^2)\beta+(\beta')^2=(\beta'_*)^2,\\
(\gamma')^2=(\gamma'_*) ^2.\label{g}
\end{eqnarray}
Here, according to the meaning of local energy balance equation, we
consider short gravitational waves (\ref{WKB}), so we can neglect
the squares of the PGW amplitudes in comparison with the squares of
their derivatives with respect to the retarded time. Thus, according
to (\ref{g}):
\begin{equation}  \label{66}
\gamma_*(u) = \gamma(u),
\end{equation}
i.e., in the linear approximation a weak gravitational waves with
polarization ${\bf e}_{\times}$ does not interact with a magnetized
plasma. This coincides with the conclusion of the paper
\cite{IgnKhu86}.

Thus, in the linear approximation the PGW with $\mathbf{e}_\times$
polarization passes through a magnetoactive plasma without
absorption, and the energy balance equation takes the form obtained
in \cite{IgnGor97}. Further conclusions are similar to the case of
propagation of the PGW with only one polarization $\mathbf{e}_+$.\\
If $\alpha^2 \gg 1$ the Eq. (\ref{eq_T_temp2}) can be written in the
form (see also \cite{Ign96}):
\begin{equation}  \label{68}
\xi^2 V(q) + {\dot q}^2 = {\dot q}_*^2\,,
\end{equation}
where $q = \beta/\beta_0$, the dot denotes differentiation with
respect to dimensionless time variable $s$:
\begin{equation} \label{69}
s = \sqrt{2} \omega u,
\end{equation}
($\omega$ - the PGW frequency), $V(q)$ - potential function which in
a weak PGW becomes:
\begin{equation}  \label{70}
V(q) = \Delta^{- 4g_\perp}(q) - 1,
\end{equation}
where $\xi^2$ is so-called {\it the first parameter of GMSW}
\cite{Ign96}:
\begin{equation}  \label{71}
\xi^2 = \frac{\stackrel{0}{H}\ \!\!\!^{2}}{4 \beta^2_0 \omega^2}.
\end{equation}
Let us introduce the new dimensionless parameter:
\begin{equation}  \label{72}
\Upsilon = 2\alpha^2\beta_0
\end{equation}
- ({\it the second GMSW parameter}) and rewrite (\ref{65}) in a weak
PGW as:
\begin{equation}  \label{73}
\Delta(q(s)) =  1 - 2 \alpha^2 \beta_0 q (s)= 1 - \Upsilon q(s).
\end{equation}
It leads from (\ref{73}):
\begin{equation}  \label{74}
{\dot q} = - \frac{\dot{\Delta}(q)}{\Upsilon}.
\end{equation}
To analyze the system behavior, let us suppose that the moment $s =
0$ corresponds to the front edge of a GW, while:
\begin{equation}  \label{75}
\beta_*\approx \beta_0(1-\cos(s))\Rightarrow q_*\approx 1-\cos(s).
\end{equation}
According to (\ref{73})-(\ref{75}) the system starts with negative
value of the governing function derivative and with function value
equal to 1:
\begin{equation}\label{76}
\begin{array}{l}
\dot{\Delta}(s)\approx -\Upsilon\sin s \approx -\Upsilon s ;\\
\\
\Delta(s)\approx 1-\Upsilon(1-\cos s)\approx 1-\Upsilon {\displaystyle \frac{s^2}{2}};\\
\end{array}
 \quad (s\to +0).
\end{equation}
The energy balance equation (\ref{68}) according to (\ref{70}),
(\ref{74}), (\ref{75}) becomes:
\begin{equation}  \label{77}
\dot{\Delta}^2 + \xi^2 \Upsilon^2 \Bigl[\Delta^{- 4g_\perp} - 1
\Bigr] = \Upsilon^2 \sin^2(s).
\end{equation}
The minimum value of the governing function at $s=\pi/2$ is equal
to:
\begin{equation}  \label{78}
\Delta_{min} = \left( \frac{1}{\xi^2} + 1\right)^{- \gamma_\perp},
\end{equation}
where:
\begin{equation} \label{79}
\gamma_\perp = \frac{1}{4 g_\perp} = \frac{1 -
k_\perp}{4}\Rightarrow \frac{1}{8}\leq \gamma_\perp \leq
\frac{1}{4}.
\end{equation}
The maximum accessible density of a magnetic energy is
\begin{equation}  \label{80}
\left(\frac{H^2}{8\pi}\right)_{max} = \frac{\stackrel{0}{H}\
\!\!\!^{2}}{8\pi} \sqrt{1 + \frac{1}{\xi^2}}
\end{equation}
and it does not depend on a plasma equation of state. Also the
plasma velocity in a GMSW does not depend on an equation of state.
And the maximum plasma energy density without magnetic field depends
on the exponent of plasma anisotropy:
\begin{equation}  \label{81}
\varepsilon_{max} = \stackrel{0}{\varepsilon} \left(1 +
\frac{1}{\xi^2}\right)^{\frac{1}{4} (1 + k_\perp)}
\end{equation}
It is maximum for ultrarelativistic plasma with zero parallel
pressure.

\section{Conclusion}
Thus, the generalization of the results of
\cite{Ign95}-\cite{2Ign96} in the case of gravitational wave with
two polarizations has been obtained and has been shown that in the
linear approximation the polarization $\mathbf{e}_\times$ does not
interact with a magnetized plasma. This fact is a justification for
applicability of the previously obtained results for the case of
arbitrarily polarized gravitational wave.

\end{document}